\documentclass{cernyrep}
\usepackage{texnames}
\usepackage[T1]{fontenc}

\usepackage{varwidth}
\usepackage{xcolor}

\begin{document}
\title{Particle Physics Instrumentation}

\author{I. Wingerter-Seez}

\institute{LAPP, CNRS, Paris, France, and Universit\'e Savoie Mont Blanc, Chamb\'ery, France}

\maketitle 

\begin{abstract}
  This reports summarizes the three lectures on particle physics instrumentation given
  during the AEPSHEP school in November 2014 at Puri-India. The lectures were
  intended to give an overview of the interaction of particles with matter and
  basic particle detection principles in the context of large detector systems like the
  Large Hadron Collider.
\end{abstract}


\section{Introduction}
\label{sec:introduction}

This report gives a very brief overview of the basis of particle detection and identification
in the context of high-energy physics (HEP). It is organized into three main parts.
A very simplified description of the interaction of particles with matter is given in
Sections~\ref{sec:basics} and~\ref{sec:interactions};  an overview of the development of electromagnetic (EM) and
hadronic showers is given in Section~\ref{sec:showers}. Sections~\ref{sec:transfer} and~\ref{sec:calorimeters}
are devoted to the principle of operation of gas- and solid-state detectors and calorimetry.
Finally, Sections~\ref{sec:pid} and~\ref{sec:detectors} give a rapid overview of the HEP detectors and a few examples.

\section{Basics of particle detection}
\label{sec:basics}

Experimental particle physics is based on high precision detectors and methods.
Particle detection exploits the characteristic interaction with matter of a few well-known particles.
The Standard Model of particle physics is based on 12 elementary fermion particles (six leptons, six quarks),
four types of spin-1 bosons (photon, W, Z, and gluon) and the spin-0 Higgs boson.
All the known particles are combinations
of the elementary fermions.
Most of the known particles are unstable hadrons which decay before reaching
any sensitive detector.
Among the few hundreds of the known particles, eight of them are the most frequently used for detection:
electron, muon, photon, charged pion, charged kaon, neutral kaon, proton and neutron.

\begin{figure}
\centering\includegraphics[width=.6\linewidth]{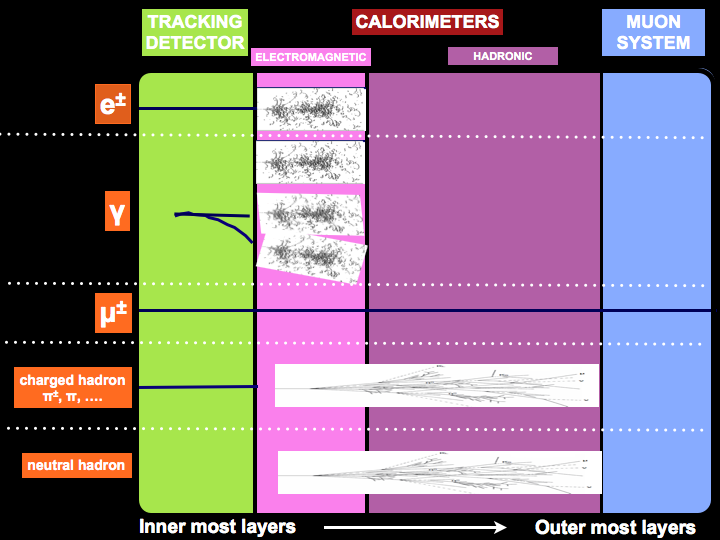}
\caption{Schematic representation of the passage of one electron, photon (unconverted and converted), $\mu^{\pm}$,
  charged hadron, and neutral hadron in a typical HEP detector built from a tracking detector, a calorimeter, and a muon detector.}
\label{fig:schematicDetector}
\end{figure}

The interaction of these particles with matter constitutes the key to detection and identification.
Figure~\ref{fig:schematicDetector} schematically illustrates the results of particle interactions
in a typical HEP detector.
The difference in mass, charge, and type of interaction constitute the means to their identification. The electron leaves a track in the
tracking detector and creates an EM shower in the calorimeter. The photon may either traverse the tracking detector
and only interact in the calorimeter, initiating an EM shower or it may convert to a pair ${\rm e}^{+}{\rm e}^{-}$ in the
tracking detector matter. The muon, which has a mass 200 times larger than the electron, traverses the entire detector,
leaving a track in the central and the muon tracking systems. Charged hadrons such as $\pi^{\pm}$ and $\rm K^{\pm}$ protons leave a track in the
tracking detector and deposit their energy in the calorimeters. The neutron and the neutral kaon $\rm K^{0}$ do not leave a track and produce a hadron
shower in the calorimeter. A neutrino traverses the entire detector without interacting but its presence can be detected via energy balance.

The role of a particle detector is to detect the passage of a particle, localize its position, measure its momentum or energy,
identify its nature, and measure its arrival time. Detection happens through particle energy loss in the traversed material. The detector
converts this energy loss to a detectable signal which is collected and interpreted.

\section{Interaction of particles with matter}
\label{sec:interactions}

The EM interaction of charged particles with matter constitutes the essence of particle detection.
Four main components of EM interaction can be identified: interaction with atomic electrons,
interaction with the atomic nucleus, and two long-range collective effects, Cherenkov and transition radiations.
Interactions with atomic electrons leads to ionization and excitation, and interactions with the nucleus lead to
Compton scattering, bremsstrahlung, and pair production (for photons). Hadrons are sensitive to the nuclear force and
therefore also obey the nuclear interaction.

This section highlights a few prominent characteristics of particle interaction with matter.
For a complete treatment and a list of references, see Ref.~\cite{bib:pdg}.

\subsection{Ionization and excitation}
\label{subsec:ionisation}

\begin{figure}
\centering\includegraphics[width=.7\linewidth]{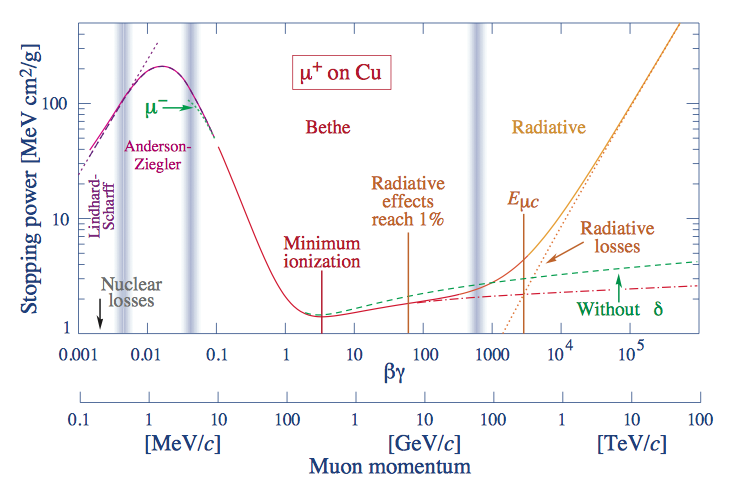}
\caption{Stopping power $-\langle\frac{\textrm{d}E}{\textrm{d}x}\rangle$ for positively charged muons in copper as a function of $\beta\gamma$ over nine
orders of magnitude in momentum.}
\label{fig:dedxmuonscopper}
\end{figure}

A heavy ($m \gg m_{\mathrm{e}}$) charged particle with $0.01 < \beta\gamma < 1000$,
passing through an atom will interact via the Coulomb force with the atomic electrons and the nucleus.
Because of the large mass difference ($2m_{\mathrm{p}}/m_{\mathrm{e}}$) between the atomic electrons and the nucleus,
the energy transfer to the atomic electrons dominates (typically by a factor of 4000).
The distribution of average energy transfer per unit distance (also called stopping power)
of positively charged muons impinging on copper is presented in Fig.~\ref{fig:dedxmuonscopper}~\cite{bib:pdg}.


\subsubsection{Energy loss for heavy particles}
\label{subsubsec:heavy}

The average energy transfer, for incoming heavy particles ($m \gg m_{\mathrm{e}}$) with $0.01 < \beta\gamma < 100$, is well described
by the Bethe formula which is reproduced in Eq.~(\ref{eq:Bethe})~\cite{bib:pdg} (terms for this are defined in Table~\ref{tab:BetheBloch})
\begin{equation}
  - \frac{\textrm{d}E}{\textrm{d}x} = K z^2 \frac{Z}{A} \frac{1}{\beta^2} \left[ \frac{1}{2} \ln{\frac{2 m_e c^2 \beta^2 \gamma^2 W_{\textrm{max}}}{I^2}} -
\beta^2 - \frac{\delta(\beta\gamma)}{2}\right] \ \ \ \left[ \UMeV\Ug^{-1}\Ucm^{2}\right]~.
\label{eq:Bethe}
\end{equation}


\begin{table}[h]
\caption{Bethe--Bloche formula terms}
\label{tab:BetheBloch}
\centering
\begin{tabular}{ll}
\hline \hline
$\phantom{0}$\\[-8pt]
$\frac{\textrm{d}E}{\textrm{d}x}$ & Average energy loss in unit of $\UMeV\Ug^{-1}\Ucm^{2}$  \\[3pt]
$K=4\pi N_{\mathrm{A}}r_{\mathrm{e}}^{2}m_{\mathrm{e}}c^{2} =$ 0.307 $\UMeV\Ug^{-1}\Ucm^{2}$ &\\
$W_{\textrm{max}}=2 m_{\ e} c^{2} \beta^{2}\gamma^{2}/(1+2\gamma m_{\mathrm{e}}/M + (m_{\mathrm{e}}/M)^{2}) $ & Maximum energy transfer in a single collision  \\
$z$ & Charge of the incident particle \\
$M$ & Mass of the incident particle \\
$Z$ & Charge number of the medium \\
$A$ & Atomic mass of the medium \\
$I$ & Mean excitation energy of the medium \\
  $\delta$& Density correction \\
  & (transverse extension of the electric field)\\
$N_{\mathrm{A}}=6.022\times 10^{23}$ & Avogadro's number \\
$r_{\mathrm{e}}=e^{2}/4\pi\epsilon_ {0}m_{\mathrm{e}}c^{2}\ $  = \Unit{2.8}{fm} & Classical electron radius \\
$m_{\mathrm{e}}$  =\ \Unit{511}{keV} & Electron mass \\
$\beta=v/c $ & Velocity \\
$\gamma=(1-\beta^{2})^{-1/2} $ & Lorenz factor \\
\hline\hline
\end{tabular}
\end{table}

The energy loss in a given material is first-order independent of the mass of the incoming particle: Eq.~(\ref{eq:Bethe})
and the curve of Fig.~\ref{fig:dedxmuonscopper} can therefore be considered universal.
The energy loss is proportional to the square of the incoming particle charge: for instance,
a helium nucleus deposits four times more energy than a proton.
From knowing the energy loss and ionization energy of a material,
it is possible to compute the number of electron--ion pairs created
along the path of the traversing particle. For a muon at minimum ionization energy traversing \Unit{1}{cm} of copper,
the number of electron--ion pairs is $\simeq 13\times 10^6/7.7 \simeq 2\times10^{6}$ with the copper ionization energy being \Unit{7.7}{eV}.

The energy loss by incoming particles leads to two effects, depending on the distance to the atomic electrons. If the distance is large,
the transferred energy will not be large enough for the electron to be extracted from the atom, and the atomic electron will go into an excited
state and emit photons. If the distance is smaller, the transferred energy can be above the binding energy, the electron will be freed,
and the atom ionized.
The photons resulting from the de-excitation of the atoms and the ionization electrons and ions are used to generate signals that
can be readout by the detector.

The understanding of energy loss of heavy particles (with $m\gg m_{\mathrm{e}}$) can be analysed in four main regimes.
\begin{enumerate}

\item The minimum ionization at $\beta\gamma\simeq$~{{3--4}} is typical and more or less universal for materials: 
  $-\frac{\textrm{d}E}{\textrm{d}x} \simeq \text{1--2} \ \UMeV \Ug^{-1} \Ucm^{2}$. As an example, for copper which has a density of
  \Unit{8.94}{g cm}$^{-3}$, a particle at minimum ionization deposits ${\simeq}\Unit{13}{MeV cm}^{-1}$.

\item For {{$\beta\gamma <$ 3--4}}, the energy loss decreases as the momentum increases, with a dependency of ${\simeq}\beta^{-2}$ as slower
particles feel the electric force of atomic electrons for a longer time.

\item For $\beta\gamma{{ > 4}}$, when the particle velocity approaches the speed of light, the decrease in energy loss should reach a minimum.
 However, the relativistic effect induces the increase of the transverse electric field, with a dependency in $\ln\gamma$, and the interaction
cross-section increases. This effect is called the relativistic rise.

\item Some further corrections to the simplified Bethe formula are necessary
to account for effects from density at high $\gamma$ values and for effects
at very low values ({{$\beta\gamma <$ 0.1 -- 1}}) where the particle velocity is close to the orbital velocity of electrons.
\end{enumerate}

As the energy loss is a function of the velocity, measuring the incoming particle momentum inside a magnetic field makes it possible to identify the particle by relating the momentum and the energy deposition, typically in the low momentum
region where the deposited energy is large.

\begin{figure}
\center\includegraphics[width=1\linewidth]{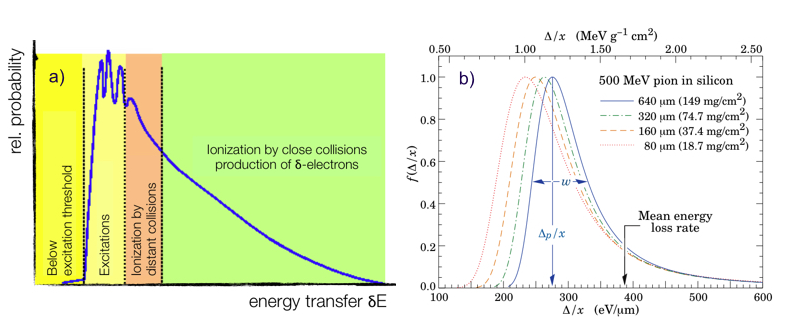}
\caption{(a) Schematic representation of the probability of energy transfer. (b) Energy loss probability described by
Landau functions for incoming \Unit{500}{MeV} charged pions in various thickness of silicon.}
\label{fig:energytransferlandau}
\end{figure}

The Bethe formula describes the mean energy deposited by a high-mass particle in a medium. The energy
deposition is a statistical process which can be described by the succession of energy loss in a material of thickness $\Delta x$,
with total energy deposition $\Delta E = \sum_{n=1}^{N}\delta E_{n}$ following the probability density function presented in
Fig.~\ref{fig:energytransferlandau}(a). Energy loss is  well described by a Landau distribution as illustrated
in Fig.~\ref{fig:energytransferlandau}(b).

\subsubsection{Energy loss for electrons and positrons}
\label{subsubsec:electrons}
The Bethe formula given in Eq.~(\ref{eq:Bethe}) is valid for heavy particles and needs to be modified to describe the energy loss of incoming electrons and positrons because the mass of the incoming particle and its target are the same.
The formula for electrons is presented as
\begin{equation}
  - \frac{\textrm{d}E}{\textrm{d}x} = \frac{1}{2} K \frac{Z}{A} \frac{1}{\beta^2} \left[ \ln{\frac{m_e c^2 \beta^2 \gamma^2 \left( m_{\mathrm{e}} c^{2} (\gamma - 1)/2 \right) }{2 I^2}} + F(\gamma ) \right] \ \ \ \left[ \UMeV \Ug^{-1} \Ucm^{2}\right]~.
\label{eq:betheblochelectrons}
\end{equation}
For positrons, this formula is different again because of the difference between electrons and positrons and to account
for the annihilation cross-section at low energies.

\subsection{Multiple scattering, bremsstrahlung and pair production}
\label{subsec:multiple}

\subsubsection{Multiple scattering}
\label{subsubsec:multiple}
The Coulomb interaction of an incoming particle with the atomic nuclei of the detector material
results in the deflection of the particle, which is called multiple scattering. The statistical sum
of many such small scattering angles results in a Gaussian angular distribution with a width,
$\theta_{0}$ given by

\begin{equation}
\theta_{0} = \frac{\Unit{13.6}{MeV}}{\beta c p} z \sqrt{x/X_{0}}\left[ 1 + 0.038\ln{(x/X_{0})}\right]~,
\label{eq:multiple}
\end{equation}

where $x$ is the distance traversed and $X_0$ is the radiation length which is introduced in Section~\ref{subsubsec:electrons2}, see
Eq.~(\ref{eq:xzero}).

Multiple scattering is an intrinsic limiting factor for tracking detectors as it induces an irreducible contribution
to the resolution. For example, the standard deviation for the multiple scattering of a \Unit{1}{GeV} incoming pion,
traversing \Unit{300}{$\mu$m} of silicon ($X_0=\Unit{9.4} {cm}$), is $\theta_{0}=\Unit{0.8}{mrad}$, corresponding
to a distance of \Unit{80}{$\mu$m} after \Unit{10}{cm} which is significantly larger than the typical resolution of
a silicon strip detector.

As shown in Eq.~(\ref{eq:multiple}), $\theta_{0}$ is inversely proportional to the particle velocity and momentum, and loss of momentum resolution from multiple scattering is greater for low-energy particles. The standard deviation is also
proportional to the square root of the material thickness in units of radiation length. Reducing the material thickness will lead to a reduction of the contribution the multiple scattering.
Tracking devices therefore favour thin material with a small radiation length, i.e. with a low atomic number (see Eq.~(\ref{eq:xzero})), such as
beryllium, used for beam pipes, carbon fibre and aluminum, used for support structures, and thin silicon or gas detectors.


\subsubsection{Interactions of high-energy electrons with matter}
\label{subsubsec:electrons2}
\begin{figure}[h]
\center\includegraphics[width=.5\linewidth]{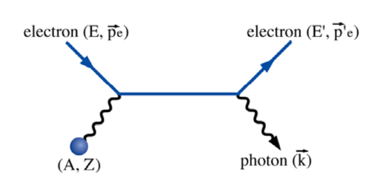}
\caption{The incoming electron (E, $\vec{p_{\mathrm{e}}}$) interacts with the nucleus (A, Z) of the traversed matter; the emitted photon
carries a momentum $\vec{k}$ and the outgoing electron (E$'$, $\vec{p_{\mathrm{e}}}'$).}
\label{fig:brem}
\end{figure}

The interaction of intermediate-energy electrons with matter is described in this section.
For incoming electrons and positrons at higher energies, the EM interaction with the nucleus is dominant.
The deflection of the charged particle by the nuclei results in acceleration and emission of EM radiation.
This effect, called bremsstrahlung, plays a key role in calorimetry measurements.
The bremsstrahlung process is represented in Fig.~\ref{fig:brem}

The spectrum of photons with energy $k$, radiated by an electron traversing a thin slab of material
has the following characteristic bremsstrahlung spectrum (dominantly in $\frac{1}{k}$) expressed
as a function of $y=k/E$~\cite{bib:DanielEDIT}:
\begin{equation}
  \frac{\textrm{d}\sigma}{\textrm{d}k} = 4 \alpha Z(Z+1) r_{\mathrm{e}}^{2} \ln{\left( 183Z^{-1/3} \right) }\left( \frac{4}{3}-\frac{3}{4}y + y^{2}\right)
  \times\frac{1}{k}~,
\label{eq:brem}
\end{equation}
where $r_{\mathrm{e}}^{2}$ is the classical radius of the electron.

The term $Z(Z+1)$ of Eq.~(\ref{eq:brem}) reflects that the bremsstrahlung results from coupling of the initial electron to the
EM field of the nucleus, augmented by a direct contribution of the atomic electrons ($Z^{2}$ replaced by $Z(Z+1)$).
The logarithmic term $\ln{(183 Z^{-1/3})}$ shows that the atomic electrons screen the nucleus field.

For a given energy $E$, the average energy lost by bremsstrahlung, $\textrm{d}E$, in a thin slab of material of thickness, $\textrm{d}x$, is obtained by integrating over $y$.
\begin{equation}
  \frac{\textrm{d}E}{\textrm{d}x} = 4 \alpha N_{\mathrm{A}} z^{2}Z^{2} \left( \frac{1}{4\pi\epsilon_{0}} \frac{e^{2}}{mc^{2}}\right) E \ln{\left( 183Z^{-1/3} \right) } \propto \frac{E}{m^{2}}~.
  \label{eq:breminteg}
\end{equation}

As shown in Eq.~(\ref{eq:breminteg}), the energy loss is proportional to $\frac{1}{m^2}$ and is therefore
mostly relevant for electrons. For a given energy, as $m_{\mu}\approx 200 \times m_{\mathrm{e}}$, the energy loss for
a muon is $\approx$ 40\,000 times smaller than for an electron. In other words, the average energy loss by bremsstrahlung
dominates ionization for muons with energy above
\Unit{400}{GeV}.

Following Eq.~(\ref{eq:breminteg}) the radiation length $X_{0}$ can be defined as the distance after which the incident
electron has radiated $(1-\frac{1}{\rm e})$ = 63\% of its incident energy and one can then write

\begin{equation*}
E(x)=E_{0} {\rm e}^{-x/X_{0}} \ \ \ \text{where}\ x\ \text{is the depth in the block of matter and}
\end{equation*}

  \begin{equation}
  \frac{\textrm{d}E}{\textrm{d}x} = \frac{E}{X_{0}}\ \ \ \text{where}\  X_{0} = 4\alpha N_{\mathrm{A}} Z^{2} r_{\mathrm{e}}^{2} \ln{183Z^{-1/3}}~.
  \label{eq:xzero}
\end{equation}

The critical energy  $E_{\mathrm{c}}$ (or $\epsilon_{0}$), at which
the average energy loss by ionization equals the
average energy loss by ionization, constitutes a useful quantity to describe the
development of EM showers. Approximation of $E_c$, for gas and liquid or solid, is given as
\begin{equation}
  E_{\mathrm{c}}^{\textrm{gas}} = \frac{\Unit{710}{MeV}}{Z+0.92}\ \ \ \text{and}\ \ \ E_{\mathrm{c}}^{\textrm{sol/liq}} = \frac{\Unit{610}{MeV}}{Z+1.24}~.
  \label{eq:critical}
\end{equation}

\begin{figure}[h]
\center\includegraphics[width=.75\linewidth]{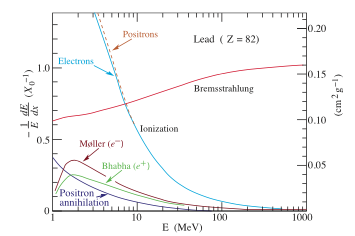}
\caption{Fractional energy loss per radiation length as a function
of electron or positron energy in lead}
\label{fig:TotalEnergyLossElectrons}
\end{figure}

Figure~\ref{fig:TotalEnergyLossElectrons}~\cite{bib:pdg} shows the fractional energy loss per radiation length as a function
of electron or positron energy.
Contributions from low-energy processes such as Moller and Bhabha scattering and ${\rm e}^{+}{\rm e}^{-}$ annihilation are shown.


\subsubsection{Interactions of photons with matter}
\label{subsubsec:photons}

\begin{figure}[h]
\center\includegraphics[width=.5\linewidth]{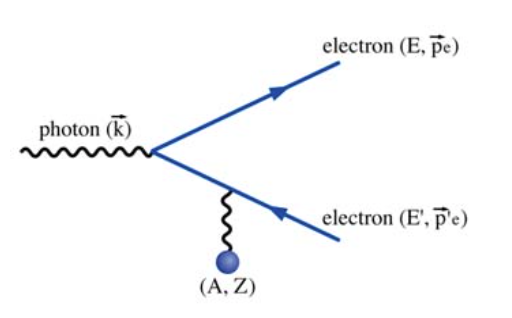}
\caption{The incoming photon ($\vec{k}$) interacts with the EM field of the nucleus (A,\,Z): an
electron ($\mathrm{E},\vec{p_{\mathrm{e}}}$), positron ($\mathrm{E}^\prime, \vec{p_{\mathrm{e}}}^\prime$) pair is created.}
\label{fig:pair}
\end{figure}

For high-energy photons for which $E> 2m_{\mathrm{e}} c^{2}$, pair creation is the dominant process as represented in Fig.~\ref{fig:pair}.
Similarly to the bremsstrahlung process for electrons, the pair creation results from the EM
interaction between the incoming photon and the field of the atom nucleus.
The pair creation cross-section is given as
\begin{equation}
\frac{\textrm{d}\sigma}{\textrm{d}x} = \frac{A}{X_{0}N_{\mathrm{A}}}\times\left( 1-\frac{4}{3}x(1-x)\right)~,
  \label{eq:pair}
\end{equation}
where $x=\frac{E}{k}$ is the fraction of the energy of the incoming photon carried by the produced electron.
The total pair production cross-section is given as
\begin{equation}
  \sigma_{\textrm{pair}} = \frac{7}{9}\frac{A}{X_{0}N_{\mathrm{A}}}=\frac{7}{9} 4 \alpha Z(Z+1)  r_{\mathrm{e}}^{2} \ln{(183Z^{-1/3})}~.
  \label{eq:pairinteg}
\end{equation}
The dominant part $Z^{2}$ is due to the interaction with the nucleus; atomic electrons contribute proportionally to $Z$.
The electron and positron are collinear as the energy of recoil of the nucleus is small (${\simeq} m_{\mathrm{e}}c^{2}$).

In addition to pair creation, photons interact in several ways.
\begin{itemize}
\item{\bf{Photo electric effect}}: for low-energy photons, the atomic electrons are not free; therefore,
the cross-section presents discontinuities whenever the photon energy crosses the electron binding energy.
This process is strongly $Z$ dependent ($\frac{Z^{5}}{E^{3.5}}$) and is dominant at very low energies.
\item{\bf{Compton scattering}}: scattering of the incoming on one atomic electron. The cross-section varies like $Z\frac{\ln{E}}{E}$.
\end{itemize}

\begin{figure}[h]
\center\includegraphics[width=.9\linewidth]{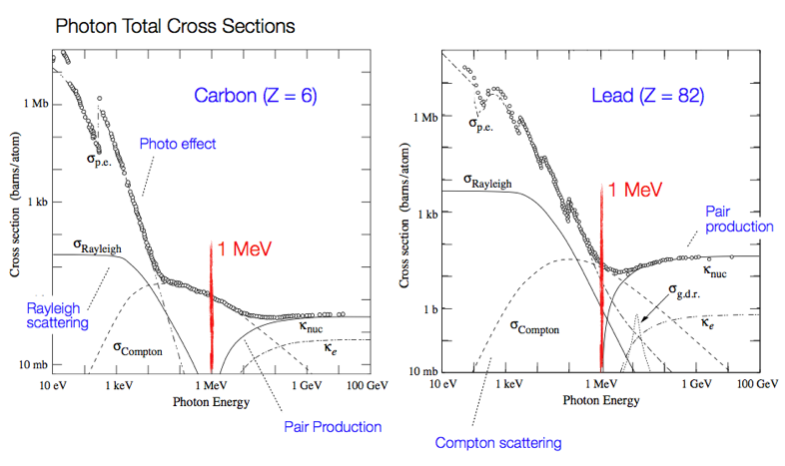}
\caption{Total cross-section
of photons impinging on carbon (left) and on lead (right)}
\label{fig:TotalCrossSectionPhotons}
\end{figure}

Photo-electrons and scattered electrons are emitted isotropically whereas electrons produced by pair creation
are emitted in the direction of the incoming photon.
Figure~\ref{fig:TotalCrossSectionPhotons} from Ref.~\cite{bib:pdg} presents the total cross-section
of photons impinging on carbon (left) and on lead (right).


High-energy electrons, positrons, and photons ($E > 100 \UMeV$) impinging on a block of material
interact dominantly with the atom nucleus via bremsstrahlung and pair-creation processes. These two processes
dominate the development of EM showers.

\subsection{Cherenkov radiation}
\label{subsec:cherenkov}
Charged particles passing through material at velocities larger than the speed of light in the material produce
an EM shock-wave that materializes as EM radiation in the visible and ultraviolet range,
the so-called Cherenkov radiation. With $n$ being the refractive index of the material, the speed of light in the material is $c/n$.
The fact that a particle does or does not emit Cherenkov radiation can then be used to apply a
threshold to its velocity. The radiation is emitted at a characteristic angle with respect to the particle direction
 $\Theta_{\mathrm{c}}=\frac{c}{nv}$. Measuring the angle of the emitted light allows measurement of the particle velocity.

\subsection{Transition radiation}
\label{subsec:tr}

Transition radiation is emitted when a charged particle crosses the boundary between two materials of different permittivities.
The probability of emission is proportional to the Lorentz factor $\gamma$ of the particle. It is only appreciable for ultra relativistic particles so it is mainly use to distinguish electrons from hadrons.
As an example, a particle with $\gamma=1000$ has a probability of about 1\% to emit one photon at the transition between two materials but,
by including many transition layers in the form of sheets, foam, or fibres, one can multiply the effect. The
energy of the emitted photons is in the \Unit{}{keV} range.

\section{EM and hadronic showers}
\label{sec:showers}

\subsection{Electromagnetic showers}
\label{subsec:emshowers}

One important consequence of the bremsstrahlung and pair-creation processes is the development of
EM showers. EM calorimeters are therefore designed so that
the shower development is contained and the deposited energy is collected.



For electrons, positrons, or photons of high energy (typically $E > \Unit{100}{MeV}$), showers result from cascading effects. Electrons undergo the bremsstrahlung process and emit photons and
photons create a pair of electrons and positrons. This cascade continues until the emitted electrons are below
the critical energy ($E_{\mathrm{c}}$ or $\epsilon_{0}$). The number of ionization electrons or photons emitted by excited atoms is
proportional to the energy of the incoming particle.

\begin{figure}[h]
\center\includegraphics[width=1\linewidth]{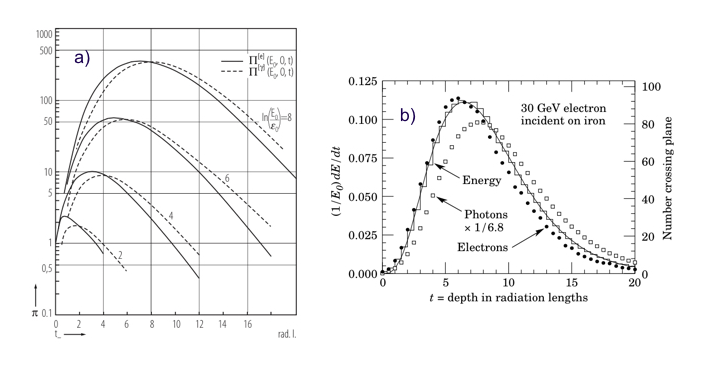}
\caption{(a) Number of electrons in electron and photon induced showers, for four energies, as a function
  of the depth in units of radiation length~\cite{bib:DanielEDIT}.
  (b) Shower profile for 60 GeV incoming electrons in iron as a function of the depth $t$ in unit of $X_{0}$~\cite{bib:pdg}.}
\label{fig:ShowerDevelopment}
\end{figure}

Figure~\ref{fig:ShowerDevelopment}(a) shows the number of electrons, as a function of the depth in units of $X_{0}$,
in EM showers induced by electrons and photons of various energies (from Ref.~\cite{bib:FournierFabjan}).
Figure~\ref{fig:ShowerDevelopment}(b) shows the longitudinal energy loss profile for electrons and photons, comparing
measurements and simulation (from Ref.~\cite{bib:pdg}).


A description of the shower development has been proposed by Rossi~\cite{bib:Rossi} by computing the number
of electrons plus positrons at a given energy and depth and similarly the number of photons.
Accounting for the processes of bremsstrahlung, Compton scattering, ionization, and pair production, the authors
have proposed a model which describes the shower development. The total track length (TTL) multiplied
by the critical energy ($E_{\mathrm{c}}$ or $\epsilon_{0}$) describes the energy transferred to the calorimeter medium
by $\frac{\textrm{d}E}{\textrm{d}x}$, which constitutes the source of the calorimeter signal.

\begin{figure}[h]
\center\includegraphics[width=.7\linewidth]{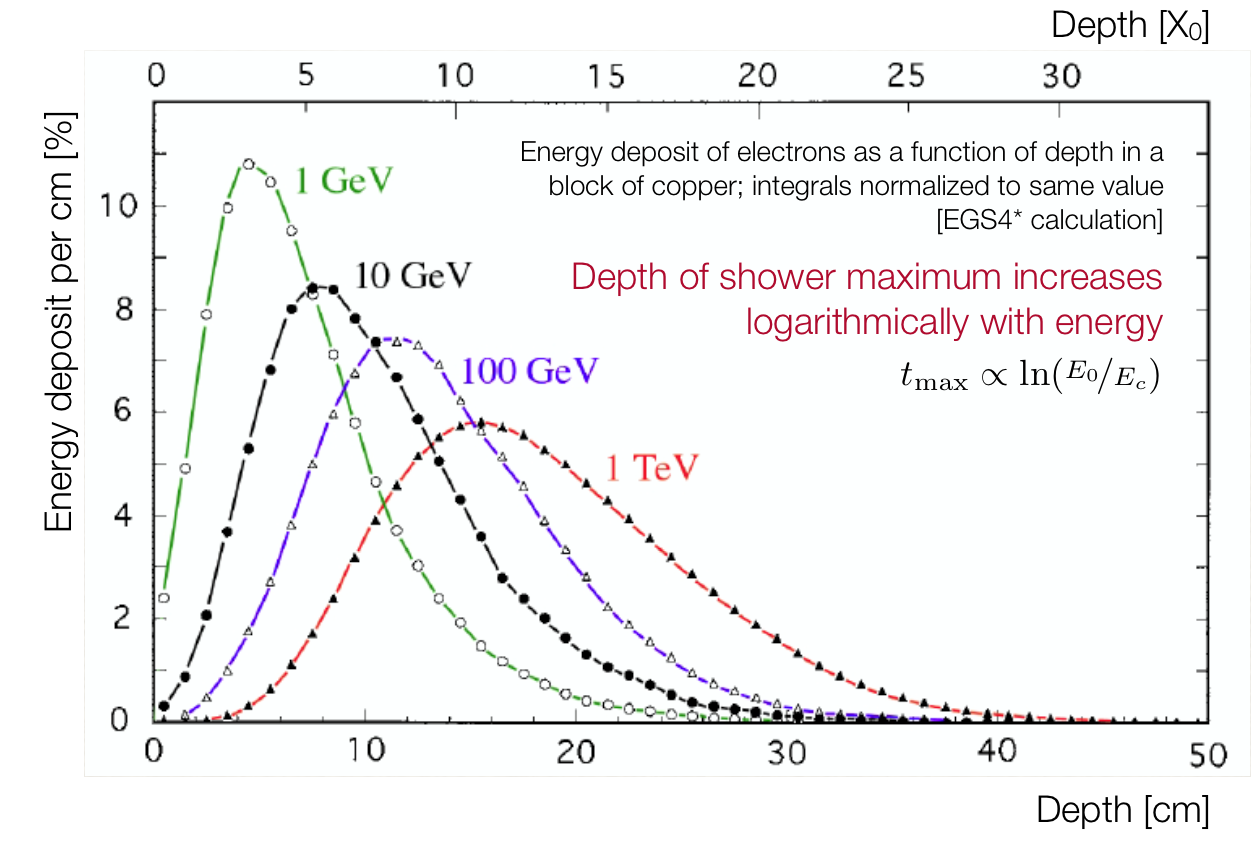}
\caption{Energy deposit of electrons with energies between \Unit{1}{GeV} and \Unit{1}{TeV}
as a function of the depth in a block of copper.}
\label{fig:ShowerProfile}
\end{figure}

In such a model, the number of segment tracks increases as the depth, $t$, in units of radiation length as
\begin{equation}
  N(t)=2^{t}
\end{equation}
and the average energy of each particle decreases as
\begin{equation}
  E(t)=E_{0}/2^{t}~,
\end{equation}
until $E(t)$ reaches the critical
energy. For $E(t)<E_{\mathrm{c}}$, ionization and excitation become dominant.
In this model, the number of tracks is maximum at
 \begin{equation}
  t_{\textrm{max}}=\ln{\frac{E_{0}}{E_{\mathrm{c}}}}/\ln{2}~,
\end{equation}
 which has important implications in the context of detector design: that the shower depth varies
 as the logarithm of the particle energy. The shape and the energy dependence of the shower profile
 are presented in Fig.~\ref{fig:ShowerProfile}.

The TTL is therefore $E_{0}\times X_{0}/E_{\mathrm{c}}$.

The higher the incident energy, the higher the TTL, and the better the relative resolution is.
The shower development for electrons and photons differs by the shift of the start of the
shower development of typically one radiation length.

\subsection{Hadronic showers}
\label{subsec:hadshowers}

Hadrons interact in matter dominantly via the nuclear interaction.
By analogy with EM showers, the energy degradation of high-energy hadrons proceeds through an
increasing number of (mostly) strong interactions with the calorimeter material.
However, the complexity of the hadronic and nuclear processes produces a multitude of effects.
The hadronic interaction produces two classes of secondary processes.
First, energetic secondary hadrons are produced with momenta that are typically a fraction of the primary
hadron momentum, i.e. at the GeV scale. Second, in hadronic collisions with the material nuclei,
a significant part of the primary energy is diverted to nuclear processes such as excitation,
nucleon evaporation, spallation, etc., resulting in particles with characteristic nuclear energies at the MeV scale.
As an example, the energy deposition of a \Unit{5}{GeV} proton impinging on a block of lead and the scintillator
can be decomposed
as 40\% ionization, 15\% EM shower, 10\% as carried by neutrons, 15\% as photons
from nuclear de-excitation, and 29\% not detectable in the form of neutrinos or binding energy.
This leads to less precision in energy resolution with respect to EM showers.

For high-energy incoming hadrons, the hadronic cross-section is fairly independent of energy and of the hadron type.
The material dependence of the total hadronic inelastic cross-section on a material of mass A is given,
in a simple form, by

\begin{equation}
\sigma_{\textrm{inelastic}}(p,A) \simeq \sigma_{0} \times A^{0.7}\ \ \text{with}\ \ \sigma_{0}=\Unit{35}{mb}~,
  \label{eq:hadxs}
\end{equation}
where $\sigma_{0}$ is the inelastic cross-section of the proton--proton interaction.

One defines the interaction length by
\begin{equation}
\lambda_{\textrm{int}} = \frac{A}{N_{\mathrm{A}}\sigma_{\textrm{inelastic}}(p,A)} \simeq 35 A^{1/3} \Ug \Ucm^2~.
\end{equation}

\section{Energy loss transfer to detectable signals and signal collection}
\label{sec:transfer}

As presented in the preceding section, charged particles traversing matter create excited atoms, electron--ion pairs
(in a gas or a liquid) or electron--hole pairs (in solid).
This section summarizes existing techniques to exploit the photons emitted by excited atoms or the ionization.

\subsection{Excitation}
\label{subsec:excitaion}
The photons emitted by the excited atoms can be detected with photon detectors such as photomultipliers or semi-conductor
photon detectors.
The emitted photons are typically in the range from ultraviolet to visible light. They are observed in
noble gases (and even liquid), inorganic crystals, and organic scintillators. The principle is to convert the $\frac{\textrm{d}E}{\textrm{d}x}$ into visible light and detect the light with a photo-sensor.
The typical light yield of scintillators is a few per cent of the energy loss. For instance, among the $\frac{\textrm{d}E}{\textrm{d}x}=1.5\UMeV$
deposited in \Unit{1}{cm} of plastic scintillator of density $\rho=1\Ug \Ucm^{-3}$, \Unit{15}{keV} are available in the form
of emitted photons and correspond to 15\,000 photons.

The main features of photo-sensors are the sensitivity to energy, the fast time response, and the pulse shape discrimination.
The requirements are high efficiency for conversion of the excitation energy to fluorescent radiation, transparency
to the radiation to allow transmission of light, emission of light in a spectral range detectable for photo-sensors, and
a short decay time to allow a fast response.
The de-excitation time is an important parameter, in particular in the context of
high-luminosity experiments such as at the Large Hadron Collider (LHC). For instance, in the PbWO$_{4}$ crystals of the CMS EM calorimeter, 
80\% of the light is emitted in 25 ns.

Two classes of scintillators are considered.
\begin{itemize}
\item{\bf{Inorganic crystals:}} which are the substance with the largest light yield and
are typically used for precision measurements of energetic photons but are typically slow.
\item{\bf{Organic scintillators:}} typically polycyclic hydrocarbons, such as naphthalene and anthracene, which typically have  a
lower light yield but a faster response than crystals.
The light produced in the scintillator propagates to the edge where it is guided in light-guides, with
total reflection, to the detector device. In addition, the use of a wavelength shifter, converting the light to
a higher wave length, allows the photon to be transported without reflecting back to the scintillator.
\end{itemize}

The classical device used to convert these photons into electrical signals is the photo-multiplier.
A photon hits a photo-cathode, a material with a very small work function, and liberates an electron which is
then accelerated in a strong electric field to a dynode, made of a material with high secondary electron yield.
The one original electron will therefore create several electrons, which are again guided to the next dynode, and so on.
Out of one single electron one ends up with a sizable signal typically of $10^{7}$--$10^{8}$ electrons. In recent years, the use
of solid-state photo-multipliers, such as avalanche photo-diodes, vacuum photo-diodes, and silicon photo-multipliers,
has become popular as they are insensitive to magnetic fields and less expensive.

\subsection{Ionization}
\label{subsec:ionization}

By applying an electric field in the detector volume, the ionization electrons and ions are moving, which induces a
signal on metallic electrodes. These signals are then readout by appropriate readout electronics.

The noise and pre-amplifier determine whether the signal can be registered. The signal-to-noise ratio must be large: $S/N \gg 1$.
The noise is characterized by the Equivalent Noise Charge (ENC) which is the charge at the input that produces
an output signal equal to the noise. The ENC of very good amplifiers can be as low as 50~$e^{-}$, a typical value being 1000~$e^{-}$.
In order to register a signal, the registered charge must be $q\gg$~ENC, i.e., typically $q\gg\text{1000}~e^{-}$.
For a gas detector, $q\simeq80~e^{-}/\Ucm$ is too small to be detected. Solid-state detectors have 1000 times more density
and a factor of 5--10 less ionization energy. Therefore, the primary charge, in a solid-state detector, therefore reaches $10^{4}$--1$0^{5}$, which is the
same for a gas detector.


\subsubsection{Gas detectors}
\label{subsub:gas}

\begin{figure}[h]
\center\includegraphics[width=.6\linewidth]{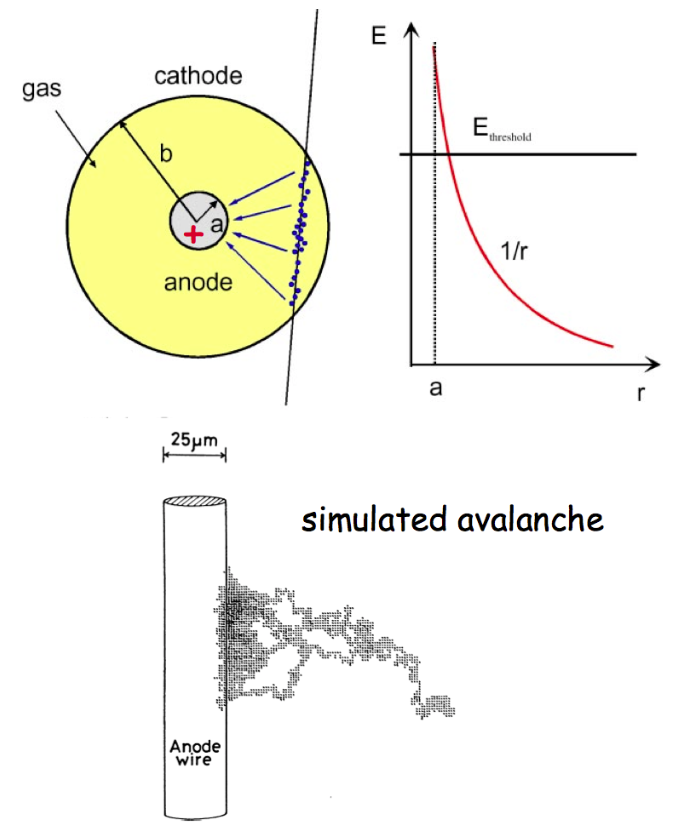}
\caption{Top-left: Schematic representation of a gas drift tube. Top-right: Electric field dependence with $r$. Bottom: Simulation
  of an ionization avalanche onto an anode wire of diameter 25 ${\mu}$m.}
\label{fig:gasamp}
\end{figure}

Gas detectors need internal amplification in order to be sensitive to a single particle.
The amplification processes and drift in an electric field are the basis of the operation of gas chambers. Ionization detectors
are generally operated in the proportional regime where an amplification of $10^{4}$ to $10^{6}$ is used.
The amplification of the signal in gas is schematically represented in Fig.~\ref{fig:gasamp}.
In the case of a cylindrical geometry, the amplification process
can be described by the electric field $E(r)$ as a function of $r$, the radial distance between the charge particle and the anode wire,
and the potential $V(r)$,
\begin{equation*}
  E(r)\propto\frac{1}{r} \ \ \ \ \text{and} \ \ \ V(r)\propto\ln{\frac{r}{a}}~.
\end{equation*}
The primary electrons drift towards the positive anode. Close to the very thin wire, due to the $1/r$ dependence, the electric
field reaches values $E>1 \UkV/\Ucm$ \\
(top-left curve of Fig.~\ref{fig:gasamp}). In between collisions with atoms, electrons gain enough energy to ionize further gas
molecules generating an exponential increase in the number of electron--ion pairs close (a few ${\mu}$m) to the wire (bottom of Fig.~\ref{fig:gasamp}).

Gas detectors are most often used as tracking detectors in order to reconstruct the charged particle trajectory and measure
its momentum from its curvature induced by the magnetic field. Criteria to obtain an optimal momentum resolution are
to have many measurement points, a large detector volume, very good single point resolution, and as little
multiple scattering as possible.

\begin{figure}[h]
\center\includegraphics[width=.6\linewidth]{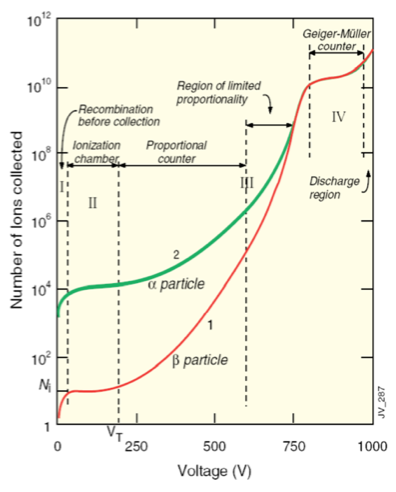}
\caption{Number of ions collected as a function of the applied voltage and definition of the operation regimes}
\label{fig:propchamb}
\end{figure}

The response of a proportional
chamber is shown in Fig.~\ref{fig:propchamb} as a function of the applied voltage. There are several distinctive regions
of the response curve: the ionization regime where the primary charge is collected, giving a flat response; the proportional regime
where the electric field is large enough to generate multiplication, with a gain up to $10^{6}$;
and the Geiger--Muller regime, where strong photon emission propagates avalanches.

The proportional mode is most often exploited with wire chambers. Wires of very small diameter, between 10 {$\mu$m} and \Unit{100}{$\mu$m},
are placed between two metallic plates a few millimeters apart. The wires are at a high voltage (a few \UkV), which results in a very
high electric field close to the wire surface. The ionization electrons move towards the thin wires, and in the strong field close
to the wires, the electrons are accelerated to energies above the gas ionization energy, which results in the production
of secondary electrons and as a consequence an electron avalanche.

The position of the primary ionization electrons can be determined by segmenting the cathodes (metal plates) into strips,
which are sensitive to the induced charge. Another way to achieve position resolution, far smaller that the separation
between the wires, is to measure the drift time of the charge with respect to a reference clock such as an accelerator clock.
For example, the precision on the track position achieved by the ATLAS muon system is of order \Unit{80}{$\mu$m}
for a distance between the wires of order \Unit{15}{mm}.

Gas chambers have been extensively used in HEP detectors. They provide many measurement points, large volume,
very good single point resolution, and as little multiple scattering as possible. They are perfectly suited for the large detector
areas at outer radii. They are not suited to the small radius of LHC experiments for instance, where the particle flow is large.
Several dedicated techniques have been employed such as multiwire proportional chambers, drift chambers, time projection chambers, streamer
tubes, and resistive plates chambers.

In the last 10--15 years, a large variety of new gas detectors have been developed in the context of particle physics instrumentation,
such as micro-pattern gas detector \'a la GEM (a gas electron multiplier) or the Micromegas (micro-mesh gas detector).

\subsubsection{Solid-state detectors}
\label{subsub:solid}

In gaseous detectors, a charged particle liberates electrons from the atoms, which are freely bouncing between the gas atoms.
An applied electric field makes the electrons and ions move, which induces signals on the metal readout electrodes.
For individual gas atoms, the electron energy levels are discrete.

In solids (crystals), the electron energy levels are in bands. Inner-shell electrons, in the lower
energy bands, are closely bound to the individual atoms and always stay with their parent atom.
However, in a crystal there are energy bands that are still bound states of the crystal, but they belong to the entire
crystal. Electrons in these bands and in the holes in the lower band can move freely around the crystal if an electric field
is applied. The lowest of these bands is called the conduction band.
If the conduction band is filled, the crystal is a conductor. If the conduction band
is empty and far away from the last filled band, the valence band, the crystal is an insulator. If the
conduction band is empty but the distance to the valence band is small, the crystal is a semi-conductor.

In order to use a semi-conductor as a detector for charged particles, the number of charge carriers in the conduction band due to
thermal excitation must be smaller than the number of charge carriers in the conduction band produced by the passage of a charged particle.
Diamond can be used for particle detection at room temperature whereas silicon and germanium must be cooled
or the free charge carriers must be eliminated by other tricks such as doping.

The diamond detector works like a solid-state ionization chamber. A diamond of a few hundred micrometres thickness is placed
between two metal electrodes and an electric field is applied. The very large electron and hole mobilities of diamond result
in very fast and short signals. Diamond is therefore both a tracking and a timing detector. Small diamond
detectors are installed as a beam condition monitor, a few centimetres from the beam pipe, in the ATLAS detector.

Silicon is the most widely used semiconductor material for particle detection.
A high-energy particle produces 33\,000 electron--hole pairs per \Ucm$^{3}$. However, at room temperature
there are $1.45\times 10^{10}$ electron--hole pairs per \Ucm$^{3}$. To be able to operate a silicon detector
at room temperature, doping is employed. Doping silicon with arsenic makes it an n-type conductor (more electrons than holes) whereas
doping silicon with boron makes it p-type conductor (more holes than electrons).
Putting an n-type and p-type conductor in contact creates a diode.

At a p--n junction, the charges are depleted and a zone that is free of charge carriers is established. By applying
a voltage, the depletion zone can be extended to the entire diode, which results in a highly insulating layer.
An ionizing particle produces free charge carriers in the diode, which drift in the electric field and
induce an electrical signal on the metal electrodes. As silicon is the most commonly used material in the electronics industry,
this constitutes a big advantage with respect to other materials.

Strip detectors are a very common application, where the detector is segmented into strips of a few 50--150~${\mu}$m pitch and
the signals are read out on the ends by wire bonding the strips to the readout electronics. The other co-ordinate can then be determined,
either by another strip detector with perpendicular orientation or by implementing perpendicular strips on the same wafer.

In the very high multiplicity region close to the collision point, a pixel detector for sizes a few tens of micrometres by
a few hundreds of micrometres can be used. The readout of a pixel module is
achieved by building the readout electronics wafer in the same geometry as the pixel layout and soldering, via bump bonding, each of the
pixels to its respective amplifier. Pixel systems of about 100 million channels are successfully operating at LHC.
The typical  vertex resolution achieved is approximately 30~$\mu$m.

Current developments in the solid-state detector domain are exploration of the possibility to integrate the detector element
and the readout electronics, as well as the application of CMOS (complementary metal-oxide semiconductor) sensors.

\section{Calorimeters}
\label{sec:calorimeters}

Calorimeters measure the energy of neutral and charged particles by complete absorption.

The principle is to measure the signal, induced by electrons and positrons with energy
below the critical energy, which is proportional to the incident energy.
Calorimeters vary by the technique used to collect the signal (sensitive material) and by the technique to induce
the shower development (passive material).
Two times two general types of calorimeters have been built: EM (Section~\ref{subsec:emcalo}) or hadronic calorimeters
(Section~\ref{subsec:hadcalo}) on one dimension
and homogeneous (Section~\ref{subsec:homocalo}) or sampling calorimeters (Section~\ref{subsec:sampcalo}) on the other dimension.

Calorimeters are a natural complement to tracking detectors as they measure the energy of both neutral and charged
particles. In calorimeters, the relative energy resolution improves with energy because it is governed by a statistical process. In contrast, relative momentum resolution degrades
with energy for tracking detectors in a magnetic field.

\subsection{Homogeneous calorimeters}
\label{subsec:homocalo}
A homogeneous calorimeter is built only from the sensitive medium. In principle, for a similar containment and signal detection
efficiency, a homogeneous calorimeter gives the best energy resolution because sampling
calorimeters are limited by {\em{sampling fluctuation}}.

The calorimeter energy resolution is determined by fluctuations, such as shower fluctuations, photo-electron
statistics, and shower leakage, and instrumental effects such as noise.
Accounting for these limitations, the relative energy resolution of a calorimeter can be written as
\begin{equation}
  \frac{\sigma(E)}{E} = \frac{a}{\sqrt{E}} \oplus \frac{b}{E} \oplus c\%
  \label{eq:caloresol}~,
\end{equation}
where $a$ is called the statistical or sampling term, $b$ the noise term, and $c$ the constant term.

For an ideal (homogeneous) calorimeter without leakage, the energy resolution is limited
only by statistical fluctuations of the number $N$ of shower particles, i.e.
\begin{equation}
  \frac{\sigma(E)}{E} \propto \frac{\sigma(N)}{N} \approx \frac{\sqrt{N}}{N} = \frac{1}{\sqrt{N}} \ \ \ \text{with} \ \ \ N=\frac{E}{W}
  \label{eq:calofluct}~,
\end{equation}
with $E$ being the energy of the incoming particle and $W$ the mean energy required to produce a {\em{signal quantum}}.
For instance $W\approx$\Unit{3.6}{eV} for a silicon detector, \Unit{30}{eV} for gas detectors and \Unit{100}{eV} for a plastic
scintillator.
The formulation of the relative energy resolution of Eq.~(\ref{eq:calofluct}) needs to be corrected to account
for correlations between fluctuations; the correction factor, $F$, is called the Fano factor~\cite{bib:FournierFabjan}:
\begin{equation}
  \frac{\sigma(E)}{E} \propto \sqrt{\frac{FW}{E}}~.
  \label{eq:calofluctFano}
\end{equation}

Homogeneous calorimeters are based on three main primary signal collection: scintillation light (PbWO$_{4}$, BGO, BaF$_{2}$),
Cherenkov light (lead glass), and ionization signal (in noble gases such as argon, krypton, and xenon).

The CMS EM calorimeter is built of ${\approx}$70\,000 lead tungsten crystal.
The energy resolution, for the barrel part of the CMS EM calorimeter~\cite{bib:CMSEresol},
reaches 1.1\% for a non-converted photon with
$E_{\perp}^{\gamma} \simeq 60 \UGeV$, and is about 1.5\% for electrons
with $E_{\perp}^{\mathrm{e}} \simeq 45 \UGeV$ from the Z$^{0}$ decay with low bremsstrahlung.
Because of inter-calibration, the measured constant term varies between 0.3\% and 0.5\% for the barrel
part of the calorimeter and between 1\% and 1.5\% in the end-cap, depending on the pseudo-rapidity.

The NA48 EM calorimeter is a homogeneous liquid krypton calorimeter.
The energy resolution for photons has been measured to be
\begin{equation*}
  \frac{\sigma_{\mathrm{e}}}{E} = \frac{(3.2\pm 0.2)\%}{\sqrt(E)} \oplus \frac{0.09\pm 0.01}{E} \oplus (0.42\pm0.05)\%~,
\end{equation*}
and the energy linearity is better than $\approx$~0.1\% for electrons in the energy range {5--100} {GeV}.
The linearity was measured using the $\frac{E}{p}$ technique, where the energy, $E$, measured by the calorimeter
is compared to the particle momentum measured from the spectrometer.

\subsection{Sampling calorimeters}
\label{subsec:sampcalo}

A sampling calorimeter consists of plates of dense, passive material alternating with layers of sensitive material.
For EM showers, passive materials with low critical energy (thus high $Z$) are used,
thus maximizing the number of electrons and positrons in a shower to be sampled by the active layers.
In practice, lead is most frequently used.
The thickness, $t$, of the passive layers (in units of $X_{0}$) determines the sampling frequency, i.e.
the number of times a high-energy electron or photon shower is {\em{sampled}}.
Intuitively, the thinner the passive layer (i.e. the higher the sampling frequency),
the better the resolution should be.
The thickness, $u$, of the active layer (in units of $X_{0}$) is usually characterized by the sampling fraction
${f}_{S}$ defined as
\begin{equation}
f_{S} = \frac{u \times \frac{\mathrm{d}E}{\mathrm{d}x}(\text{active})}{u \times \frac{\mathrm{d}E}{\mathrm{d}x}(\text{active}) + t\times \frac{\mathrm{d}E}{\mathrm{d}x}(\text{passive})}  \end{equation}
where $u, t$ are in $\Ug\,\Ucm^{-2}$ and $\frac{\mathrm{d}E}{\mathrm{d}x}$ is in $\UMeV\,\Ug^{-1}\,\Ucm^{2}$.

This {\em{sampling}} of the energy results in a loss of information and hence in additional sampling fluctuations.
An approximation~\cite{bib:FournierFabjan} for these fluctuations in EM calorimeters
can be derived using the TTL (see Section~\ref{subsec:emshowers}) of a shower, initiated by an electron or photon of energy $E$.

The signal is approximated by the number $N_{x}$ of e$^{+}$ or e$^{-}$
traversing the active signal planes, spaced by a distance ($t+u$).

This number $N_{x}$ of crossings is
\begin{equation}
N_{x} = \frac{TTL}{t+u} = \frac{E}{E_{\mathrm{c}}}\frac{1}{t+u}~.
\end{equation}
Assuming statistical independence of the
crossings, the fluctuations in $N_{x}$ represent the {\em{sampling fluctuations}} $\sigma(E)_{\textrm{samp}}$ ,
\begin{equation}
 \frac{\sigma(E)_{\textrm{samp}}}{E} = \frac{\sigma(N_{x})}{N_{x}} = \frac{1}{\sqrt{N_{x}}} = \frac{a}{\sqrt{E[\UGeV]}}~.
\end{equation}

\begin{figure}[h]
\center\includegraphics[width=.9\linewidth]{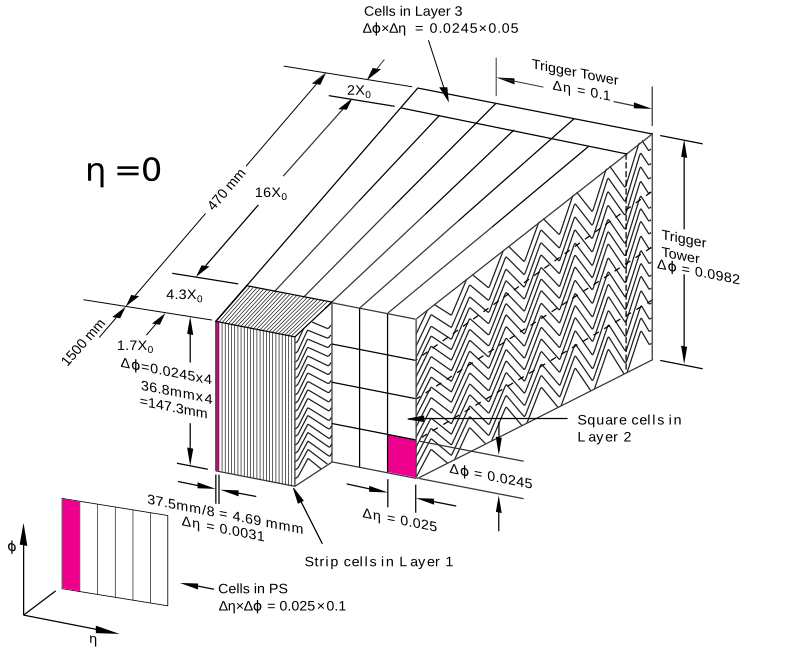}
\caption{Schematic representation of the ATLAS liquid argon calorimeter }
\label{fig:EMLAr}
\end{figure}

As an example, Fig.~\ref{fig:EMLAr} shows a schematic view of the barrel section of the ATLAS
EM calorimeter which
is built with lead plates (e.g. 1.53\Ucm\ thick for the central part), interleaved with copper electrodes to build a \Unit{6.4}{m}
cylinder of \Unit{2.8}{m} inner diameter and \Unit{4}{m} outer diameter.
The lead absorbers and the electrodes are folded with an accordion shape which allows signal
collection on the front and back faces of the calorimeter. This has two main advantages: a fast
signal collection and the absence of gaps between cells, leading to a complete coverage in the azimuthal direction.
Signal pads are drawn on the electrodes to measure the particle position and sample the shower
development.
The cylinder is housed in a cryostat filled with liquid argon (LAr), and
the lead constitutes the passive material and LAr the active material. The ionization electrons
drift towards the electrodes through which a high voltage is applied. A current is induced which is
driven to preamplifiers located outside of the cryostat.

The energy resolution, for the barrel part of the ATLAS EM calorimeter~\cite{bib:ATLASEresol},
is  1.5\% for non-converted photons with
$E_{\perp}^{\gamma} \simeq 60 \UGeV$, and is about 1.5\% for electrons
with $E_{\perp}^{\mathrm{e}} \simeq 45 \UGeV$ from the Z$^{0}$ decay.
The measured constant term varies between 0.7\% and 1\% for the barrel
part of the calorimeter and between 1\% and 2.5\% in the end-cap, depending on the pseudo-rapidity.

\subsection{EM calorimeters}
\label{subsec:emcalo}

EM showers develop as described in Section~\ref{subsec:emshowers}. The number of particles
in the shower increases until the average energy of the produced particle is below the critical energy when
no more particles can be produced. The particles in the shower will ionize the medium or undergo Compton scattering.
The lateral development of the shower is mainly governed by the electrons that do not radiate but have
enough energy to travel far away from the axis.

The typical relative energy resolution for sampling EM calorimeters at the LHC is
\begin{equation}
\frac{\sigma(E)}{E} = \frac{10\%}{\sqrt{E[\UGeV]}} \oplus \frac{100~\UMeV}{E[\UGeV]} \oplus 0.5\text{--}1\%~.
\end{equation}

\subsection{Hadronic calorimeters}
\label{subsec:hadcalo}

The development of hadronic showers is more complex than the development of EM particles (see Section~\ref{subsec:hadshowers}).
The fraction of detectable energy from hadronic showers is smaller than for EM showers, leading
to an intrinsically worse relative energy resolution for hadrons than for electrons and photons.
In order to contain high-energy hadrons, hadronic calorimeters need to be larger than EM calorimeters.
As an example, the interaction length $\lambda_{\textrm{int}}$, which describes the typical size of one nuclear interaction, is
\Unit{17}{cm} in iron when the radiation length $X_{0}$ is \Unit{1.7}{cm}.
In addition, as hadronic showers extend deeper and wider than EM showers, the granularity
of a hadronic calorimeter is coarser than for an EM calorimeter.
Hadron calorimeters are mainly sampling calorimeters, with iron or copper as the passive material and a scintillator
as the active material.
The typical relative energy resolution of hadronic calorimeters for the LHC is
\begin{equation}
\frac{\sigma(E)}{E}  = \frac{\text{50--100\%}}{\sqrt{E[\UGeV]}}~.
  \end{equation}

\section{Particle identification}
\label{sec:pid}

Particle identification is typically the result of the combination of several observations from
various instruments. Only a few examples are given here.

\begin{figure}[h]
\center\includegraphics[width=.7\linewidth]{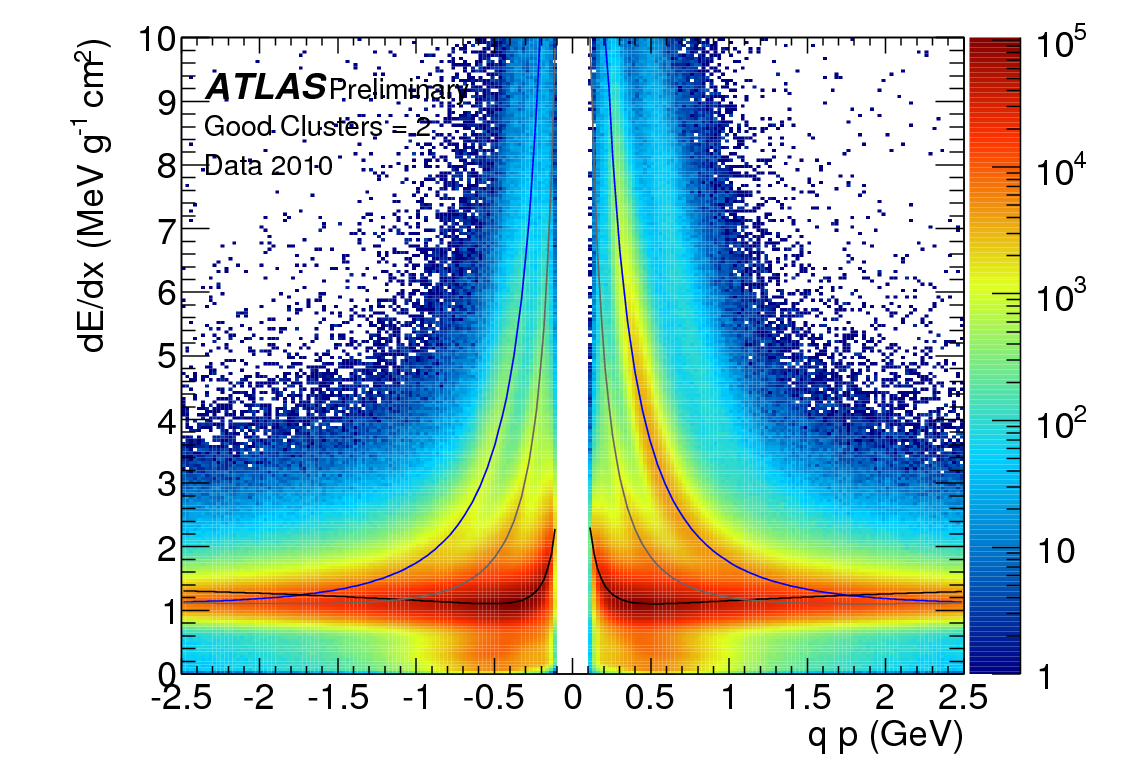}
\caption{Bi-dimensional distribution of $\frac{\textrm{d}E}{\textrm{d}x}$ and momentum for ATLAS 2010 data.
  The distributions of the most probable value for the fitted probability density functions of pions (black), kaons (grey),
  and protons (blue), in different track categories, are superimposed.}
\label{fig:dedx}
\end{figure}

By combining the energy loss along a charged track, which is a function of the velocity $\beta$,
with the momentum measurement from the curvature in the magnetic field, one can extract the particle
mass, providing the momentum is small enough. This is illustrated by Fig.~\ref{fig:dedx} from Ref.~\cite{bib:dedxATLAS}.

\begin{figure}[h]
\center\includegraphics[width=.6\linewidth]{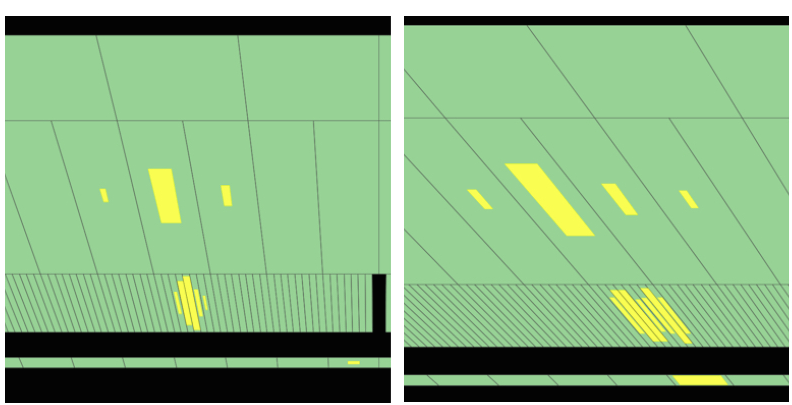}
\caption{Energy depositions in the successive layers of the ATLAS EM calorimeter of a candidate photon on the left
  and a candidate $\pi^{0}$ on the right.}
\label{fig:photonpizero}
\end{figure}

The ATLAS EM calorimeter segmentation allows identification of single photons from $\pi^{0}$ decays to two photons
by its ability to separate single from double showers in the thinly segmented first layer. The is illustrated
in Fig.~\ref{fig:photonpizero}. The resulting rejection power  of the calorimeter for $\pi^{0}$ is about three for high-energy isolated photons.

As illustrated by Fig.~\ref{fig:schematicDetector}, the association of the tracks measured in the inner tracking system
with the energy deposits in the calorimeter makes it possible to distinguish between electrons and photons.
The shape of the energy deposition in the EM and hadronic calorimeters makes it possible to distinguish between electrons and hadrons.

\section{Detectors}
\label{sec:detectors}

Additional components are necessary to build and operate a complete detector.
In the previous sections, only a few topics have been discussed. Many
more would need to be included to give a fair description of the versatility,
complexity and refinement of the design and operation of a detector.
Among the missing topics are: low noise electronics, fast digital electronics,
selective and efficient trigger systems, high data flow systems, real-time
software, highly radiation tolerant components, reconstruction software, and calibration.

Revealing and measuring rare phenomena are the main goals for LHC physics. This imposes extremely
severe constraints on the four main LHC detectors: ATLAS, CMS, LHCb, and ALICE.
These four detectors are large, complex, and have very high capability.
Huge magnet systems dominate their mechanical structures.
The proton collision rate of 1~GHz produces particles and jets of TeV-scale energy. This
imposes severe demands in terms of spectrometer and calorimeter size, rate capability,
and radiation resistance. The fact that a few hundred events, out of the $10^{9}$
events produced every second, can be written to disk necessitates highly complex
online event selection as a trigger.

The basic layout of LHC collider experiments is quite similar.
\begin{itemize}
\item{\bf{Tracking detector:}} charged particles are bent inside a solenoidal
  magnetic field produced by the magnet surrounding the tracking system. The charged particle
  momentum is reconstructed from the association of hits collected in the
  sensitive tracking detectors along the track path.
  \begin{itemize}
  \item{\bf{Vertex detector:}} close to the interaction point,
  there are several layers of pixel detectors which allow the
  position of charged particle vertices to be measured to a few tens of micrometres. This
  also allows short-lived B and D mesons to be identified.
\item{\bf{Spectrometer:}} to follow the track curvature a succession of sensitive layers
  (mainly silicon strips) are positioned around the interaction region and typically
  up to \Unit{1}{m} in radius and a few metres along the beam line.
  The CMS tracking detector is built from silicon sensors only, whereas the other experiments use silicon at low radius and a gas detector further along the track path.
  \end{itemize}
\item{\bf{Calorimeters:}}
  the tracking detector is surrounded by the EM and hadronic calorimeters,
  which measure the energy of electrons, photons, hadrons, and jets by absorbing
  them. The hermiticity of the calorimeters in the transverse direction and down
  to very small angles close to the beam line allows reconstruction of the total transverse
  energy deposited with a high-performance resolution thereby allowing reconstruction of the
  transverse energy carried away by neutrinos or any non-interacting particle.
  The LHC calorimeters are highly segmented in order reconstruct the position of
  neutral particles, to separate electrons from hadrons, and to be less sensitive
  to the energy pile-up from simultaneous proton--proton interactions.
\item{\bf{Muons systems:}}
  muons, which deposit very little energy in the calorimeters and are therefore not
  stopped, are measured at very large radii by dedicated muons systems.
\end{itemize}

The sequence of the vertex detector, spectrometer, calorimetry, and muon detector is the
classic basic geometry that underlies most collider and fixed-target experiments.
There are many other types of detectors, from very small to very large arrays
of telescopes, for example. Each detector has its own specialism
such as neutrino detection or detection of very high-energy gamma rays from astrophysical sources.
These are only a few examples from a large variety of existing detector systems.
However, it is important to remember that there are only a few basic principles of particle
interaction with matter that underlie these different detectors.

\section*{Acknowledgements}
I have profited enormously, in the preparation of the lectures and of these proceedings, from
the material prepared by Hans Christian Schultz-Coulon~\cite{bib:HansChristian},
Marco Delmastro~\cite{bib:MarcoESIPAP}, Daniel Fournier~\cite{bib:DanielEDIT}, and
Werner Riegler~\cite{bib:WeinerRaigler}.

\end{document}